# Band gap and band alignment prediction of nitride based semiconductors using machine learning

Yang Huang[1,2], Changyou Yu[3], Weiguang Chen[4], Yuhuai Liu[3,5], Chong Li[1], Chunyao Niu[1], Fei Wang[1,*], Yu Jia[1,6,*]

[1] International Laboratory for Quantum Functional Materials of Henan, School of Physics and Engineering, Zhengzhou University, Zhengzhou 450001, China

[2] Materials Science and Engineering Program, University of California San Diego, La Jolla, California 92093, USA

[3] National Center for International Joint Research of Electronic Materials and Systems, School of Information Engineering, Zhengzhou University, Zhengzhou, Zhengzhou 450001, China

[4] Quantum Materials Research Center, College of Physics and Electronic Engineering, Zhengzhou Normal University, Zhengzhou 450044, China

[5] Institute of Materials and Systems for Sustainability, Nagoya University, Nagoya 464-8603, Japan

[6] Key Laboratory for Special Functional Material of Ministry of Education, and School of Materials Science, Henan University, Kaifeng 475004, China

## Abstract

Nitride has been drawing much attention due to its wide range of applications in optoelectronics and remains plenty of room for materials design and discovery. Here, a large set of nitrides have been designed, with their band gap and alignment being studied by first-principles calculations combined with machine learning. Band gap and band offset against wurtzite GaN accurately calculated by the combination of screened hybrid functional of HSE and DFT-PBE were used to train and test machine learning models. After comparison among different techniques of machine learning, when elemental properties are taken as features, support vector regression (SVR) with radial kernel performs best for predicting both band gap and band offset with prediction root mean square error (RMSE) of 0.298 eV and 0.183 eV, respectively. The former is within HSE calculation uncertainty and the latter is small enough to provide reliable predictions. Additionally, abstractwhen band gap calculated by DFT-PBE was added into the feature space, band gap prediction RMSE decreases to 0.099 eV. Through a feature engineering algorithm, elemental feature space based band gap prediction RMSE further drops by around 0.005 eV and the relative importance of elemental properties for band gap prediction was revealed. Finally, band gap and band offset of all designed nitrides were predicted and two trends were noticed that as the number of cation types increases, band gap tends to narrow down while band offset tends to go up. The predicted results will be a useful guidance for precise investigation on nitride engineering.

[*] E-mail: wfei@zzu.edu.cn; jiayu@zzu.edu.cn

# Introduction

Machine learning, a popular data mining technology that has been widely used in computer vision, speech recognition and natural language processing, has also recently been effectively used for materials research [1], specifically, in property prediction [2] and prescreening in high-throughput materials search [3, 4]. On the other hand, nitride semiconductor materials have emerged as one of the most important classes of materials in modern semiconductor industry over the past 40 years. This family of materials that traditionally consists of wurtzite III-N binary compounds such as AlN, GaN, InN and later involves II-IV compounds like $Zn(Sn,Ge)N_2$ with various forms of alloys have shown multiple significant applications in light-emitting diodes, lasers, photodetectors and photovoltaics due to a broad range of band gap values ranging from deep UV to terahertz [5]. Despite the great accomplishment it has achieved, nitride semiconductor is still relatively unexplored compared to other families of materials such as oxides and remains a broad space for materials discovery and design [6]. Nitride design is highly motivated by the great number of possible but unexplored structures with promising optoelectronic properties. Structural diversity and property uniqueness of nitride partly origins from high valence (-3) of nitrogen element which requires either metal elements with high valence or a large number of low valence metal elements in various ways of combinations in a formula unit of a nitride compound. In nitride semiconductor design, the most worth-investigating property is band gap because it is the determinate factor that affects the performance of nitride semiconductor in optoelectronic device and is thus also regarded as the most commercially significant property. It has been demonstrated that band gap exhibits satisfying tunability when alloying different metal elements or altering compositions in the nitride compound [7, 8]. In addition, nitride materials have also been used in semiconductor heterojunctions [9, 10]. In semiconductor heterojunction engineering, band offset between two connecting materials acts as the key parameter that determines junction performance such as potential barrier and mobility [11]. Therefore, designing new nitride materials through elemental and compositional modulation followed with band gap and band offset measurement or calculation should be an effective way for new nitride semiconductor discovery.

Due to the aforementioned extremely large amount of possible nitride structures, experimentally, it is difficult to fabricate and characterize the overall possible new nitride semiconductors. From a theoretical perspective, although conventional density functional theory (DFT) is relatively computationally efficient, it suffers from obvious band gap underestimation [12]. Accurate band gap calculations require advanced method such as screened hybrid functionals of Heyd-Scuseria-Ernzerhof (HSE) [13] or many body perturbation theories [14], however, they are both far more computationally expensive than DFT and not able to be applied to large materials set. For band offset calculation, although using DFT for interfacial potential alignment is accurate enough [15], typically, a superlattice that consists of hundreds of atoms needs to be built up even for a simple compound, which is also computationally expensive when applied to large materials set. A successful machine learning model is typically trained by a small subset of a large dataset and is able to predict the whole dataset within an acceptable error. In the case of band gap and band offset calculation in new nitrides, if accurate first-principles calculations are performed on a small subset of nitrides or their junctions and the results are used to train a machine learning model, it is highly likely that all nitrides in the design space can be accurately predicted. Pieces of work on band gap prediction by using machine learning methods have been reported: Zhuo *et al*. used 136 engineered elemental features and SVR model trained and tested on 3896 various forms of semiconductors for experimental band gap prediction, achieving a RMSE of 0.45 eV [16]. By using 18 features including both elemental properties and low-level DFT computational results of compounds, Lee *et al*. used SVR model on 270 binary and ternary compounds and achieved a RMSE of 0.24 eV in experimental band gap prediction [17]. Weston *el al*. trained and tested SVR model on 284 $I_2$-II-IV-$VI_4$ kesterite compounds with HSE calculated band gaps by using 12 elemental features, achieving a RMSE of 0.283 eV [18]. To the best of our knowledge, by using a machine learning approach, neither systematic work on nitride band gap prediction nor band offset prediction for bulk materials has ever been reported.

In the present paper, 16-atom constructed wurtzite nitrides in orthorhombic cell were studied and 68115 possible materials were considered based on all possible cation-nitrogen combinations in the design space. 300 out of the total 68115 materials were randomly selected and their band gap was calculated by using the hybrid functionals of HSE method and band offset against wurtzite GaN was calculated by the combination of HSE and DFT-GGA (Generalized Gradient Approximation) based on interface models. Calculated results were used to train and test machine learning models. Various machine learning models were tested and their performances were compared with each other in terms of RMSE. By using 18 accessible elemental properties as features, radial kernel SVR with an RMSE of 0.183 eV performs best for band offset prediction. For band gap prediction, radial kernel SVR is again the best model and shows an RMSE of 0.298 eV with the same 18 elemental features. Through a feature engineering, 26 elemental properties were taken as features and RMSE decreases by around 0.005 eV compared to 0.298 eV and the relative importance of elemental properties for band gap prediction was found. Our results show that the designed nitrides exist in all valuable band gap ranges and interestingly, as the number of the types of cations increases from 1 to 8, mean band gap decreases and mean band offset increases. Both the predicted values and discovered trends with cation type number will be useful as guidance for computational and experimental investigations on nitride engineering with higher precision.

## Methodology

**a. Materials design space**

Materials studied in this paper have been derived from 16-atom 2×2×2 supercell of wurtzite GaN by cation transmutations and combinations. In the design space, +2, +3 and +4 cations were considered to occupy the positions near nitrogen anions. +2 cations are from group II A: $Be^{2+}$ $Mg^{2+}$ $Ca^{2+}$ $Sr^{2+}$ $Ba^{2+}$ and group II B: $Zn^{2+}$ $Cd^{2+}$; +3 cations are from group III A: $Al^{3+}$ $Ga^{3+}$ $In^{3+}$ and group III B: $Sc^{3+}$ $Y^{3+}$; +4 cations are from group IV A: $Si^{4+}$ $Ge^{4+}$ $Sn^{4+}$ and group IV B: $Ti^{4+}$ $Zr^{4+}$ $Hf^{4+}$. With the consideration of proper size of the total materials set, occupations are divided into 3 types and rules are shown in Figure 1. All eight cation positions are entirely symmetric. Type 1: I III V VII are occupied by +2 cations while II IV VI VIII are occupied by +4 cations; type 2: all 8 cation positions are occupied by +3 cations; type 3: I III are occupied by +2 cations, II IV are occupied by +4 cations while all the rest cation positions are occupied by +3 cations. In total, 68115 different nitrides were constructed and 300 materials were randomly selected for training and testing of machine learning model. These 300 nitrides with feature space and all computational results by first-principles calculations are listed in Table S1 in supplemental information. For the nitride compounds designed in this work, due to the high electronegativity of nitrogen atom, binding energy of the system is usually high and it can be expected that the nitrides in the design space should be thermally stable. It has been checked that the formation energies of the 300 randomly selected materials are all negative. These calculated formation energies are listed in Table S1 in supplemental information. Therefore, the designed 16-atom supercell based on wurtzite GaN is reasonable in this study.

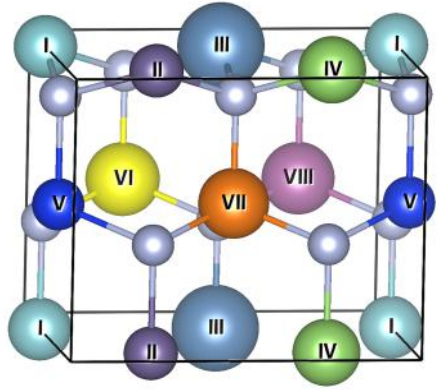

Figure 1: The nitride structure in the design space and positions of 16 ions. 8 labelled balls are cations and the other 8 unlabelled balls are nitrogen atoms.

**b. First-principles calculation**

The first-principles calculations have been performed using the plane-wave pseudopotential as implemented in the VASP code [19, 20]. The electron-core interactions are described with the frozen-core projected augmented wave pseudopotentials [21]. The generalized gradient approximation (GGA) formulated by Perdew, Burke, and Ernzerhof (PBE) as the exchange-correlation functional [22] with cut-off energy of 500 eV for basic functions was chosen in all of our calculations. A reciprocal space sampling of 6 × 5 × 6 Monkhorst-Pack mesh [23] of the Brillouin zone is used in the structural optimizations. All the structures are fully relaxed until the forces on each atom are smaller than 0.01 eV/Å with a tetrahedron method with Blöchl corrections in broadening of 0.05 eV. The screened hybrid functional of HSE with α= 0.31 rather than the typical value 0.25 [26] has been performed on PBE optimized structures for band gap calculation. A comparison test between HSE calculated bandgap based on exchange parameter of 0.31, typical value 0.25 and experimental or GW calculated band gap values for seven nitrides in the design space has indicated that 0.31 leads to more accurate bandgap values. The test result has been tabulated in Table S5 in supplemental information. For band offset calculation, 300 superlattices have been formed on the PBE optimized isolated nitrides along the (001) direction as $(XN)_n/(GaN)_n$, where n = 5 and XN represents each of the 300 nitrides we have random selected. All the energy levels in isolated materials have been calculated through HSE method and energy levels in the constructed interface models have been calculated at the DFT-PBE level.

By using DFT-PBE method, for each constructed nitride compound, energy of the compound ($E_C$) and energy of most stable elementary substance of every component element ($E_{I-VIII}$) are calculated respectively. Formation energies ($E_f$) were obtained through equation (1):

$$E_f = E_C - \sum_I^{VIII} E_i \tag{1}$$

Band offset against wurtzite GaN was calculated using Wei's core level method [25] [26] with following equation (2):

$$\Delta E_V(XN/GaN) = \Delta E_{V,C'}(XN) - \Delta E_{V,C}(GaN) + \Delta E_{C'/C}(XN/GaN) + A_V(XN) + A_V(GaN) \tag{2}$$

where $\Delta E_{V,C'}(XN) = E_V(XN) - E_{C'}(XN)$, $\Delta E_{V,C}(GaN) = E_V(GaN) - E_C(GaN)$, $E_V$ is the energy level of valence band maximum (VBM) in isolated materials, $E_C$ ($E_{C'}$) is core energy level in isolated materials, $\Delta E_{C'/C}$ is the core energy level difference between two materials at both sides of an interface model constructed and $A_V$ is the valence band deformation potential. Core level has been set to be the 1s level of nitrogen atom due to the adequate low energy which is around -370 eV. Considering the geometric similarity among constructed nitrides in the design space, for simplicity, it was assumed that valence band deformation potential of each compound when connected with wurtzite GaN was neglected, i.e., $A_V(GaN) + A_V(XN) = 0$. Consequently, the band offset between wurtzite XN and GaN can be written as $\Delta E_{V,C'}(XN) - \Delta E_{V,C}(GaN) + \Delta E_{C'/C}(XN/GaN)$. Calculated band offsets were compared with widely accepted reported results by Wei [25] in Table 1, which shows a satisfying consistence.

|         | Band offset calculated (eV) | Band offset reported (eV) |
| --- | --- | --- |
| AlN/GaN | -1.11 | -1.28 [25] |
| InN/AlN | 1.16 | 1.11 [25] |

Table 1: Computational band offset comparison between our method and Wei' s results.

**c. Machine learning**

Machine learning work in this paper was implemented in Python 2.7 code with frameworks Scikit-learn [27] for SVR and Tensorflow [28] for linear regression and neural network. Three types of machine learning models: support vector regression (SVR) with linear, polynomial and radial kernels, linear regression and neural network (NN) with single hidden layer (ANN) and two hidden layers (DNN) are used. All hyper-parameters were optimized. In order to prevent overfitting, $L_2$ regularization term was added to the loss function of NN models. According to previous work, covalent radius, electronegativity and valence of each component element are three of the most common elemental features chosen for machine learning predictions on band gap properties. For example, electronegativity, ionic radius, and row in the periodic table have been used by Weston *el al*. in the prediction of bandgap of $I_2$-II-IV-$VI_4$ kesterite compounds [18].

Elemental information including absolute value of the formal ionic charge, period in the periodic table, atomic number, atomic mass, van der Waals radius, electronegativity, and the first ionization energy have been chosen as features for bandgap predictions in binary and ternary compounds by Lee *el al*. [17]. From a view of physical intuition, covalent radius, electronegativity and valence are the most important electronic properties of an element and should be the most relevant and effective descriptors for electronic band gap and alignment predictions in this study. Therefore, these three elemental features have been chosen as an initial trial. After removing symmetrically repeated values, an 18-dimensional feature space was built and used first. Model performance was evaluated by averaged RMSE of validation set in 10-fold cross validation.

## Results & Discussion

In order to check if 300 randomly selected nitrides are adequate or not to effectively learn band gap and band offset, the Leave-out-one-cross-validation (LOOCV) RMSEs of radial kernel SVR models trained and tested on randomly selected subsets of 300 nitride samples were calculated. Band gap and band offset prediction RMSE with subset size are plotted in Figure 2 (a), (b), respectively. The curves were fitted with power functions [29] and it is shown that with 300 nitrides, the prediction capacity is adequately stable and has almost reached its limit. Therefore, a sample set with a size of 300 should be large enough for a machine learning model to learn in this work.

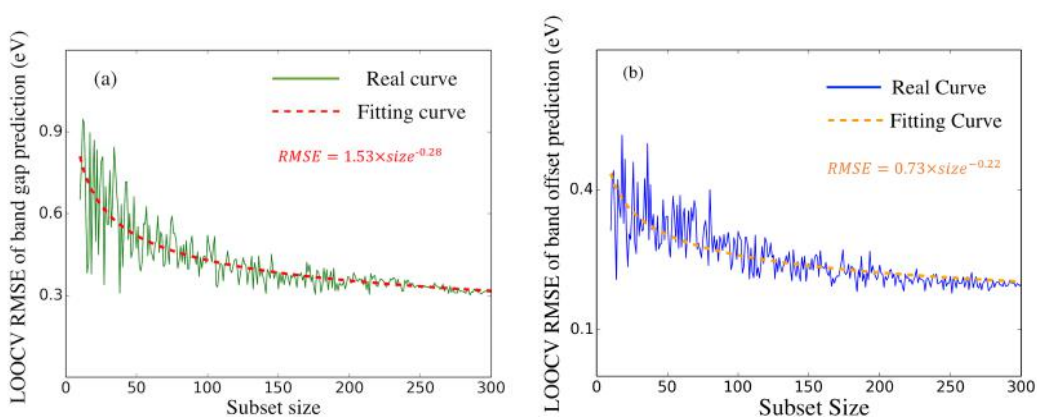

Figure 2: (a) LOOCV RMSE of band gap prediction versus subset size (green curve). Red dash curve is the fitting curve fitted by the power function shown in red. (b) LOOCV RMSE of band offset prediction versus subset size (blue curve). Orange dash curve is the fitting curve fitted by the power function shown in orange.

a. **Band offset regressor**

For band offset prediction, by using the 18-dimensional elemental feature space, RMSEs of all models are shown in Table 2. Optimized hyper-parameters are listed in Table S2 in supplemental information.

|  | Linear SVR | Poly SVR | Radial SVR | Linear regression | ANN | DNN |
|---|---|---|---|---|---|---|
| Band offset RMSE (eV) | 0.256 | 0.239 | 0.183 | 0.256 | 0.219 | 0.230 |
| Band gap RMSE (eV) | 0.412 | 0.335 | 0.298 | 0.474 | 0.385 | 0.379 |

Table 2: RMSE of band offset and band gap prediction for different machine learning models. RMSE values listed are averaged RMSE of validation set in 10-fold cross validation.

The smallest RMSE, 0.183 eV, suggests that SVR with radial kernel is the best model for band offset prediction. In order to intuitively show the accuracy of band offset prediction, predicted band offset values as a function of calculated band offset in both training and validation set are plotted in Figure 3. The excellent prediction performance in both training set and validation set indicates that the model is neither under-fitting nor over-fitting.

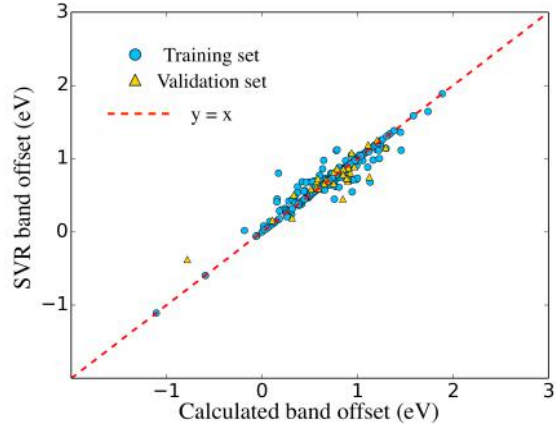

Figure 3: SVR predicted band offset versus calculated band offset. The blue circles represent training set, the gold triangles represent validation set and the red dash line is the guidance line on which prediction error is zero.

**b. Band gap regressor**

For band gap prediction, RMSEs of all optimized models trained with the elemental 18-feature space are shown in Table 2. Optimized hyper-parameters are listed in Table S2 in supplemental information.

SVR with radial kernel again performs best for band gap prediction and came up with a RMSE of 0.298 eV. From a perspective of first-principles calculations, HSE calculated band gap is sensitive to exchange parameter with empirically selected values, which gives rise to a band gap calculation uncertainty reaching up to 0.4 eV [26]. Since the RMSE of 0.298 eV is within the HSE calculation uncertainty, the model performance is satisfactory. When band gap of each nitride compound calculated by DFT-PBE method was added into the 18-dimensional feature space, a radial kernel based SVR model was trained under the new 19-dimensional feature space and validation RMSE becomes as low as 0.099 eV. A performance comparison of radial kernel SVR models with the 18-dimensional and 19-dimensional feature space is shown in Figure 4. The tremendous accuracy enhancement by introducing PBE band gap can be explained by the approximately linear relationship between PBE band gap and HSE band gap [30].

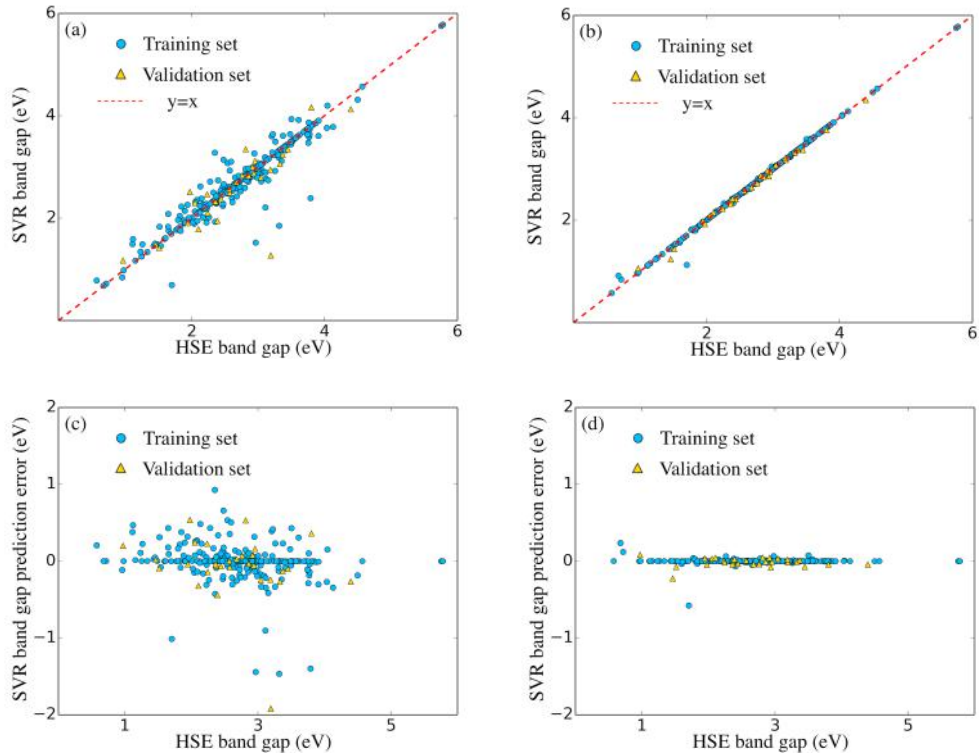

Figure 4: (a) SVR predicted band gap versus HSE calculated band gap in 18-dimensional elemental property based feature space. (b) SVR predicted band gap versus HSE calculated band gap in 19-dimensional PBE band-gap-included feature space. (c) SVR band gap prediction error (the difference between SVR predicted band gap and HSE calculated band gap) versus HSE calculated band gap in 18-dimensional elemental property based feature space. (d) SVR band gap prediction error versus HSE calculated band gap in 19-dimensional PBE-band-gap-included feature space. Blue circles represent training set and gold triangles represent validation set.

As a trial to further improve the performance of SVR with radial kernel on band gap prediction based on elemental properties, feature space expansion implemented by an elemental property-based recursive feature extraction (EPRFE) algorithm was conducted. In EPRFE, firstly, a larger feature space which includes all accessible and reportedly-band-gap-related elemental properties was built. After removing symmetrically repeating features, a new 58-dimensional feature space was established and includes 8 elemental properties: covalent radius, electronegativity, valence, atomic number, periodic number, atomic weight, first ionization energy and melting point. Secondly, models were trained and tested with feature space that is the subset of the 58-dimentional space based on all possible combinations of the 8 elemental properties and the validation RMSEs of all 255 combinations were compared. The lowest RMSEs with the corresponding number of properties selected are shown in Figure 5. Interestingly, it was found that the lowest RMSE when 3 properties are selected corresponds to covalent radius, electronegativity and valence, exactly the three properties in the original 18-dimensional elemental feature space. RMSE can be further decreased a little bit by around 0.005 eV when first ionization energies were introduced as new features, which corresponds to the case of 4 properties selected in Figure 5. Besides, in Figure 5, the lowest RMSE of 1 property corresponds to electronegativity and the lowest RMSE of 2 properties corresponds to electronegativity and covalent radius, which indicates that the relative importance of each property for SVR based band gap prediction from high to low is electronegativity, covalent radius, valence and first ionization energy. Other more complex feature engineering methods such as different orders of polynomial feature combinations with filtering method for large-scaled feature selection were tried, however, no improvement was observed on the model performance.

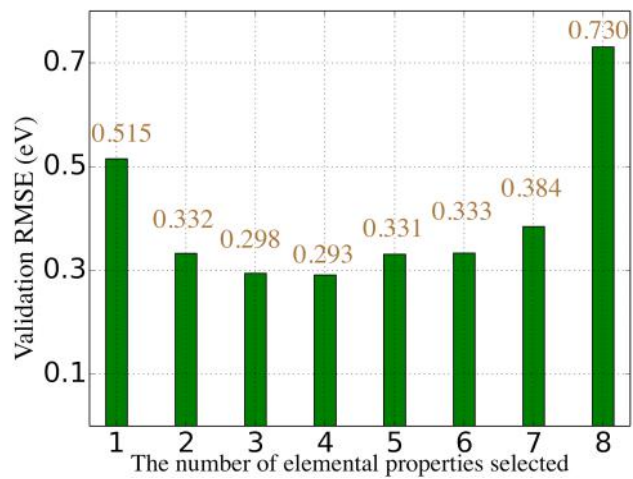

Figure 5: Lowest validation RMSEs in 10-fold cross validation with the number of elemental properties selected.

**c. Predicted results**

Band gap and band offset against wurtzite GaN of all 68115 constructed nitrides were predicted by using radial SVR with 26-dimensional (original 18 features plus 8 first ionization energies) and the original 18-dimensional feature space, respectively. Predicted band gaps and band offsets of all 68115 nitrides are listed in Table S4 in supplemental information. Predicted band gap of several nitrides that have been previously investigated are listed in Table 3 in a comparison with reported and HSE calculated results. In Table 4, some previously unexplored nitrides are listed with band gap being categorized into three application domains: infrared detector, solar cell absorber and ultraviolet LED. Overall distributions of predicted band gap and band offset with Gaussian fitting curves are shown in Figure 6. Distributions of band gap and band offset by different numbers of cation types with Gaussian fitting curves are shown in Figure 7. It was found that designed nitrides exist in all valuable band gap ranges and pretty interestingly, as the number of cation types increase, both the mean and Gaussian mean band gap tends to decrease while both the mean and Gaussian mean band offset tends to increase. Mean and Gaussian mean values with the number of cation types are listed in Table S3 in supplemental information. It is suggested that both theorists and experimentalists can make further investigations on their interested nitrides included among the predicted results in this work. Specifically, people can find their targeted materials by looking for satisfied bandgap in the band gap database predicted through the band gap regressor. When searching for materials to make heterojunctions, targeted materials can be found by screening both band gap and band offset data generated by both regressors. Furthermore, when looking for promising materials for specific device applications, various device parameters need to be taken into consideration. If device parameter regressors would have been built for targeted applications such as infrared detector, solar cell absorber and ultraviolet LED, then combined with band gap and band offset regressors, materials with potentials for excellent device performance can be found from materials database based on the three sorts of regressors.

|  | Reported Eg (eV) | Predicted Eg (Eg) | HSE calculated Eg (eV) |
|---|---|---|---|
| $AlGaN_2$ | 4.650 [7] | 4.570 | 4.569 |
| $InGaN_2$ | 1.795 [7] | 2.272 | 1.925 |
| $AlInN_2$ | 2.890 [7] | 3.445 | 2.976 |
| $ZnGeN_2$ | 3.420 [31] | 3.405 | 3.406 |
| $ZnSnN_2$ | 2.020 [31] | 2.155 | 1.566 |
| $MgGeN_2$ | 5.140 [32] | 4.857 | 4.304 |
| $MgSiN_2$ | 5.840 [32] | 5.753 | 5.755 |
| $CaSiN_2$ | 4.500 [33] | 4.701 | 5.072 |

Table 3: Comparison of predicted band gap of previously explored nitrides with reported and HSE values.

| Infrared detector | Eg (eV) | Solar cell absorber | Eg (eV) | Ultraviolet LED | Eg (eV) |
|---|---|---|---|---|---|
| $BeBaSn_2In_4N_8$ | 0.016 | $CaSnGa_2N_4$ | 1.226 | $BeMgSiTiN_4$ | 4.862 |
| $CdSnIn_2N_4$ | 0.044 | $CdSiSn_2N_4$ | 1.306 | $BeMg_3Si_2Ge_2N_8$ | 4.964 |
| $CdSnGaInN_4$ | 0.345 | $BaCdSn_2N_4$ | 1.341 | $Mg_4GeTi_3N_8$ | 5.008 |
| $SrSnIn_2N_4$ | 0.371 | $SrCdSn_2N_4$ | 1.368 | $Mg_2SiGeN_4$ | 5.020 |
| $BaSnIn_2N_4$ | 0.420 | $BaGeGa_2N_4$ | 1.483 | $BeSiAl_2N_4$ | 5.106 |
| $CdGeIn_2N_4$ | 0.554 | $CdGeGa_2N_4$ | 1.491 | $BeMgSi_2N_4$ | 5.457 |
| $CaSnIn_2N_4$ | 0.561 | $SrGeGa_2N_4$ | 1.499 | $BeMg_3Si_2TiZrN_8$ | 5.020 |
| $ZnSnIn_2N_4$ | 0.596 | $SrSnGaYN_4$ | 1.191 | $BeMg_3Si_2GeTiN_8$ | 5.056 |

Table 4: Predicted band gap of selected previously unexplored nitrides in three domains of applications, infrared detector, solar cell absorber and ultraviolet LED

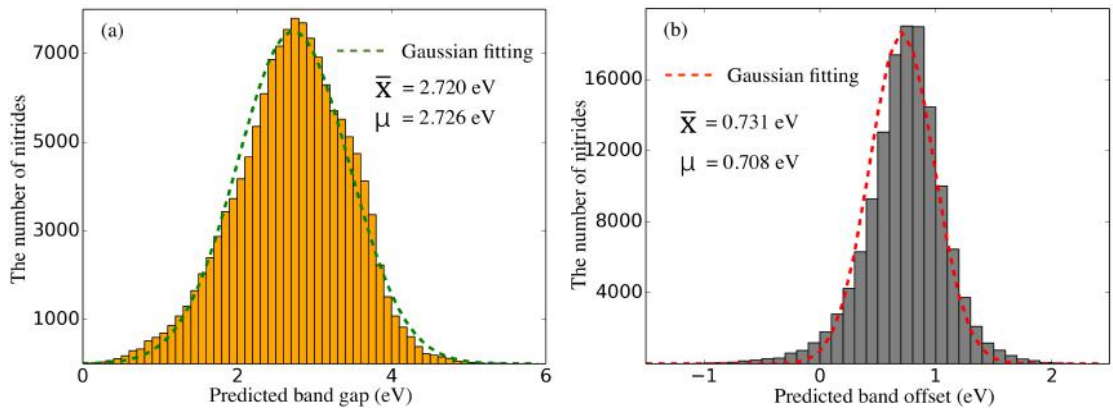

Figure 6: (a): Distribution of predicted band gaps of all designed nitrides. (b): Distribution of predicted band offset (against wurtzite GaN) of all designed nitrides. The green and red dash curve is the fitting curve fitted by Gaussian function. $\bar{x}$ is the mean value of predicted results. $\mu$ is the mean value of the Gaussian fitting curves.

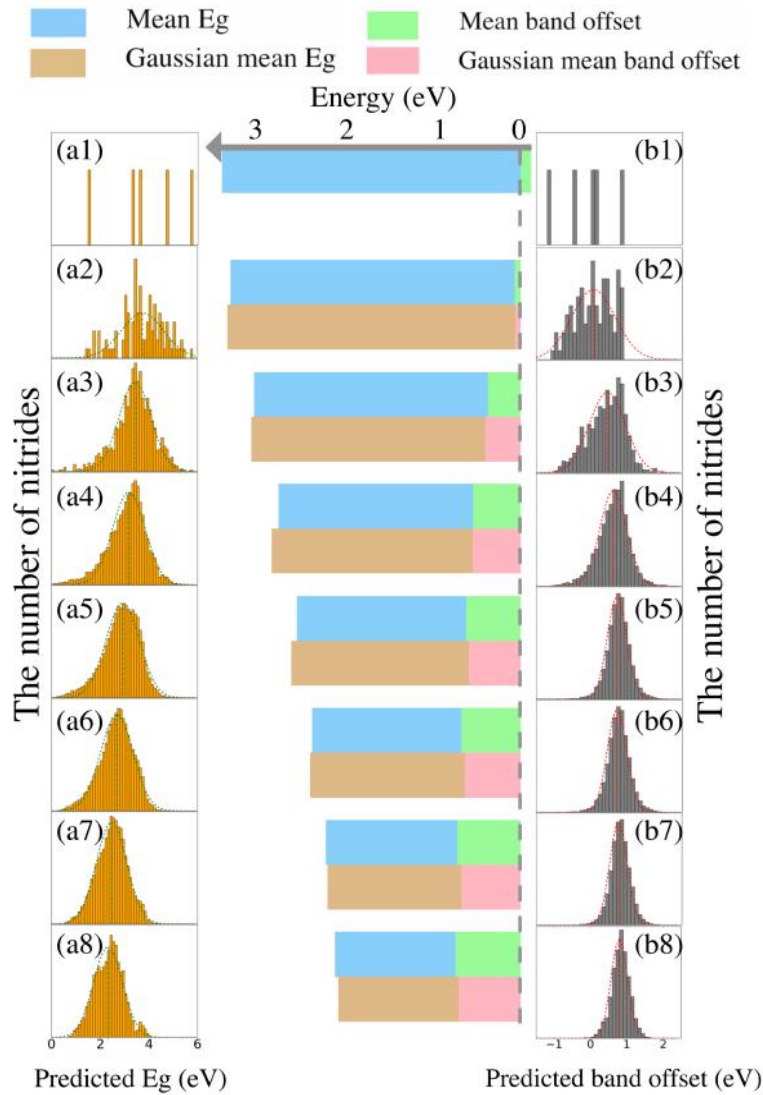

Figure 7: Left: Distribution of predicted band gap of all designed nitrides with 1-8 types ((a1) - (a8)) of cations. Right: Distribution of predicted band offset against wurtzite GaN of all designed nitrides with 1-8 types ((b1) - (b8)) of cations. Middle: Mean and Gaussian mean of predicted band gaps and band offsets versus the number of types of cations, horizontally matched with left and right figures. Gaussian fittings for one type of cation were not made due to small sample size.

## Conclusions

In this work, machine learning models trained on first-principles calculations results were utilized to successfully provide accurate predictions for band gap and band alignment of nitrides in a large design set. After model comparison, SVR with radial kernel function came up with the lowest RMSE of 0.183 eV for band offset prediction and through feature engineering, a RMSE of 0.293 eV for band gap prediction. It was found that when DFT-PBE calculated band gap was introduced into the feature space, band gap prediction RMSE could jump down to 0.099 eV. Eventually, band gap and band offset were predicted on the total 68115 nitrides in the design space and nitrides with useful band gaps and alignment were discovered. The prediction results also indicate that the more types of cations a nitride includes, the smaller band gap and larger band offset it tends to have. Along with the predicted results, further investigations can be conducted on new nitride semiconductor materials with desired applications.

**Acknowledgements**

This work was jointly supported by National Natural Science Foundation of China (Grant No. 11774078), National Key R&D Program of China (2016YFE0118400), and Outstanding Young Talent Research Fund of Zhengzhou University (Grant No. 1521317008). The calculations were performed on the high performance computational center of Zhengzhou University.

**Conflicts of interest**

There are no conflicts of interest to declare